\documentclass[]{aastex631}

\begin{document}

\title[Article Title]{Drivers of Magnetic Field Amplification at Oblique Shocks: In-Situ Observations}

\author[0000-0002-2234-5312]{Hadi Madanian}
\affiliation{Laboratory for Atmospheric and Space Physics, \\
University of Colorado Boulder, \\
Boulder, CO 800303, USA}

\author[0000-0003-2218-1909]{Imogen Gingell}
\affiliation{School of Physics and Astronomy, \\
University of Southampton \\
Southampton, SO17 1BJ, UK}

\author[0000-0002-4768-189X]{Li-Jen Chen}
\affiliation{NASA Goddard Space Flight Center, \\
Greenbelt,MD 20723, USA}

\author{Eli Monyek}
\affiliation{Astrophysical and Planetary Sciences Department, \\
University of Colorado Boulder, \\
Boulder, CO 80309, USA}

\begin{abstract}
Collisionless shocks are ubiquitous structures throughout the universe. Shock waves in space and astrophysical plasmas convert the energy of a fast-flowing plasma to other forms of energy, including thermal and magnetic energies. Plasma turbulence and high-amplitude electric and magnetic fluctuations are necessary for effective energy conversion and particle acceleration. We survey and characterize in-situ observations of reflected ions and magnetic field amplification rates at quasi-perpendicular shocks under a wide range of upstream conditions. We report magnetic amplification factors as high as 25 times the upstream magnetic field in our current dataset. Reflected ions interacting with the incoming plasma create magnetic perturbations which cause magnetic amplification in upstream and downstream regions of quasi-perpendicular shocks. Our observations show that in general magnetic amplification increases with the fraction of reflected ions, which itself increases with Mach number. Both parameters plateau once full reflection is reached. Magnetic amplification continuously increases with the inverse of the magnetization parameter of the upstream plasma. We find that the extended foot region upstream of shocks and nonlinear processes within that region are key factors for intense magnetic amplification. Our observations at non-relativistic shocks provide the first experimental evidence that below a certain magnetization threshold, the magnetic amplification efficiency at quasi-perpendicular shocks becomes comparable to that at the quasi-parallel shocks.

\end{abstract}

\section{Introduction} \label{sec:intro}

Our understanding of distant astrophysical shocks is based on remote observations of emissions and accelerated particles at these structures. Magnetic amplification at astrophysical shocks has direct impacts on the post-shock plasma temperature and associated X-ray emissions \citep{blandford_particle_1978}, and on the generation of galactic cosmic rays at supernova remnants \citep{bell_cosmic_2013}. Magnetic amplification at collisionless shocks is proportional to the amount of upstream energy \citep{sulaiman_quasiperpendicular_2015, russell_overshoots_1982}. The source of this energy, among others, can be from the shock expansion into the upstream medium compressing the upstream field \citep{donnert_magnetic_2018}, density inhomogeneities and shock-generated vorticities and turbulence \citep{fraschetti_turbulent_2013, ji_efficiency_2016}, and upstream magnetic perturbations induced by backstreaming or reflected ions (RIs) \citep{madanian_dynamics_2021, bell_turbulent_2004, bohdan_magnetic_2021}. 

Oblique quasi-perpendicular ($Q_{\perp}$) shocks, which are focused in this study, travel at large angles ($\theta_{Bn} > 45^{\circ}$) to the upstream magnetic field ($\textbf{B}_U$). At strong supercritical shocks, a fraction of the incident ions is reflected upstream to allow for heating and thermalization of already transmitted ions. Particle reflection is through a combination of the cross-shock electrostatic potential and magnetic deflection \citep{gedalin_ion_1997,madanian_direct_2021}, with the latter (former) becoming more (less) important with increasing Mach number (the ratio of the shock speed to a characteristic wave speed in the medium) \citep{leroy_structure_1982,dimmock_statistical_2012}. RIs play a key role in energy repartition across the shock. At shocks with significant particle reflection, most of the enthalpy flux downstream of the shock is carried by the suprathermal population returning RIs that have crossed the shock layer into the downstream \citep{schwartz_energy_2022}. The kinematics of RIs behind the shock front is associated with the magnetic overshoot and undershoot structures \citep{leroy_structure_1982}, and relaxation of the directly transmitted ions \citep{gedalin_kinematic_2019}. Presence of RIs upstream of the shock front leads to a modest and gradual increase in the upstream magnetic field and formation of the shock ``foot'', with roughly the size of an ion gyroradius \citep{Sckopke1983}. In the normal incidence frame, in which the upstream flow is along the normal vector to the shock surface $\hat{\textbf{\textit{n}}}$, the gyroradius of a specularly reflected proton is $r_{L}=2 m_p V_U sin(\theta_{Bn}) / e B_U$, where $m_p$ is the proton mass and $V_U$ is the upstream flow speed. A weaker magnetic field in the upstream plasma corresponds to a larger foot region. For non-planar bow shocks, the reflection point across the shock also influences the ion trajectory \citep{madanian_transient_2023, gedalin_non-locality_2023}.

RIs are a source of free energy upstream of the shock layer, making the plasma unstable to develop electrostatic and electromagnetic waves \citep{buneman_instability_1958, weibel_spontaneously_1959, fried_mechanism_1959}. Relative drift between RIs and the upstream electrons and ions leads to various forms of streaming instability and magnetic perturbations that can propagate at oblique angles to the background magnetic field and towards upstream or downstream \citep{marcowith_microphysics_2016}. These waves are steepened and nonlinearly amplified as they traverse the foot, and can modify the shock front \citep{madanian_dynamics_2021,madanian_nonstationary_2020,burgess_microstructure_2016}. In this paper, we investigate factors influencing the magnetic amplification rates at $Q_{\perp}$ shocks by analyzing the properties of RIs that are responsible for magnetic field amplification.

\section{Data and Method}\label{sec:method}

\begin{figure}[h!]
\centering
\includegraphics[width=0.4\textwidth]{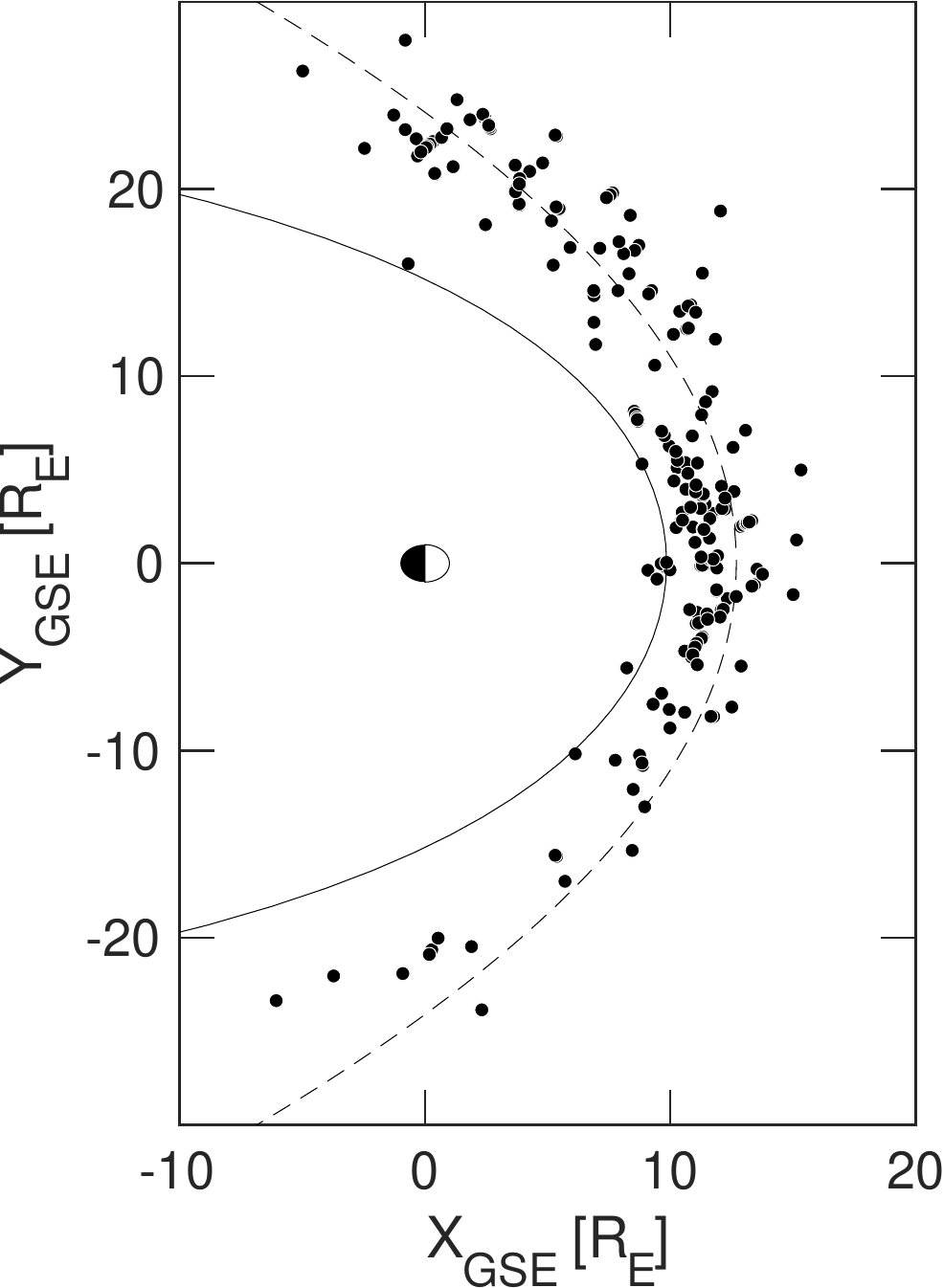}
\caption{\textbf{The distribution of the Earth's bow shock crossing events.} The events are projected on the $XY$ plane of the Geocentric solar ecliptic (GSE) coordinates, in which the $+X$ axis is towards the Sun, the $+Y$ axis points opposite to the planetary motion, and the $+Z$ axis completes the right-hand triple. Nominal bow shock (dashed line) and magnetopause (solid line) boundaries are shown for reference \citep{farris_determining_1994,shue_magnetopause_1998}.  The axes are in units of Earth radii (R\textsubscript{E}).
\label{fig:fig1}}
\end{figure}

We analyze data from the Magnetospheric Multiscale (MMS) mission \citep{burch_magnetospheric_2016}, a set of four spacecraft orbiting the Earth. During specific phases of the mission, the spacecraft trajectories extend into the solar wind and cross the Earth's bow shock, enabling high time resolution in-situ measurements of shock processes. Shock events in MMS data are surveyed and cataloged through ongoing activities and reported in previous studies \citep{lalti_database_2022}. Our analysis includes 222 $Q_{\perp}$ shock events. The distribution of these events across the nominal bow shock boundary is shown in Figure \ref{fig:fig1}.

MMS spacecraft are equipped with plasma and field instruments including the Fast Plasma Investigation (FPI) unit that measures charged particles \citep{pollock_fast_2016} and the magnetometer system \citep{russell_magnetospheric_2016}. Due to the coarse resolution at very low energies, FPI measurements may underestimate the density and overestimate the temperature of the cold solar wind beam. For each event, we determine the upstream plasma temperature $T_U$ from the solar wind monitor measurements at Lagrange point 1 shifted to the nose of the bow shock \citep{king_solar_2005}. The upstream plasma density $n_U$ is selected by comparing MMS measurements in the pristine solar wind and the shifted solar wind monitor data and selecting the higher value. We obtain $\textit{B}_U$ from MMS data in the unperturbed solar wind. The Alfvén Mach number, $M_{Alf} = v_{shock} / v_{Alf}$ is calculated using the solar wind flow speed along the shock normal direction as a proxy for the shock speed ($v_{shock}$). $v_{Alf} = B_U / \sqrt{\mu_0 \rho}$ is the Alfvén speed in which $\mu_0$ is the vacuum permeability and $\rho$ is the plasma mass density. The distribution function of various ion populations in units of energy flux $\phi(E, \Omega)$ is derived from FPI count rate measurements in field-of-view (FOV, $d\Omega$) and energy ($dE$) bins. For an ion population occupying a specific region of the phase space, the total particle flux, $\textbf{\textit{F}}=n\textbf{\textit{V}}$, can be obtained by integrating the energy flux across the occupied solid angles and energy ranges:

\begin{equation}\label{eq:flux}
\textbf{\textit{F}} = \sum \sum \frac{\phi(E, \Omega)}{E}dE d\Omega
\end{equation}

The population density is determined from:

\begin{equation}\label{eq:dens}
n = \sqrt{\frac{m}{2}} \sum \sum \frac{\phi(E, \Omega)}{E^{\frac{3}{2}}}dE d\Omega
\end{equation}

We define the momentum tensor, $\bar{\textbf{P}}=m<v>$, as:
\begin{equation}
\bar{\textbf{P}} = \sqrt{2m} \sum \sum \sqrt{E} \phi(E, \Omega) dE d\Omega
\end{equation}

\noindent which we use to obtain the pressure tensor in the frame of reference of the ion population:
\begin{equation}
\textbf{p} = \bar{\textbf{P}} - m \textbf{\textit{V}} \textbf{\textit{F}}
\end{equation}

For some bow shock crossing events, the RI population upstream of the shock is clearly separate from the solar wind ions in the FPI FOV which allows us to study RI properties by calculating plasma moments for each population. The separation is dependent on the Lorentz force factor in the upstream plasma. The two populations can be tracked and distinguished from one another typically up to the shock ramp, beyond which ions are scattered, heated, and mixed by the enhanced magnetic field. We label these events as ``RI Track''. See the Supplementary Figure \ref{fig:supp_ref_track} for an example. In addition to density and velocity (Eqs. \ref{eq:flux} and \ref{eq:dens}), the RI beam temperature ($T_{RI}$) can be obtained by taking the average of the two smallest eigenvalues of the temperature tensor, $\textbf{p}/n$. For events that a unique set of FOV bins for RIs cannot be selected, we calculate the density of the solar wind beam based on its trace in the FPI FOV and its energy range and subtract it from the total density to obtain the density of the RIs. These events are labeled as ``SW Subtract'' and an example is shown in Supplementary Figure \ref{fig:supp_sw_subt}.

The abundance of RIs decreases rapidly with distance to the shock front. To make reasonable comparisons of the ion reflection rate between various shock events, the RI density must be selected at the same distance to each shock. Given the dynamic motion of the shocks and nonstationarity effects, consistent and reliable identification of a specific position in the foot region of every shock becomes nontrivial. However, there are several indicators in the data for identifying the shock front, such as the jump in the magnetic field, heating of solar wind electrons, and deflection of the solar wind ion beam in the FPI FOV. For each event, we first determine the shock crossing time and then isolate the solar wind and RI populations and calculate the moments. We select the RI density ($n_{RI}$) at the closest point to the shock front where the moments could be determined. Roughly, half of that density is associated with freshly-reflected ions. It is worth noting that steepened waves or nonstationarity effects can appear in time series data as transient structures containing heated, compressed, and nearly isotropic plasma much like the downstream plasma. These instances are easily identified in the data and are avoided during the selection.

\section{Results}\label{sec2}

\begin{figure}[h!]
\centering
\includegraphics[width=0.8\textwidth]{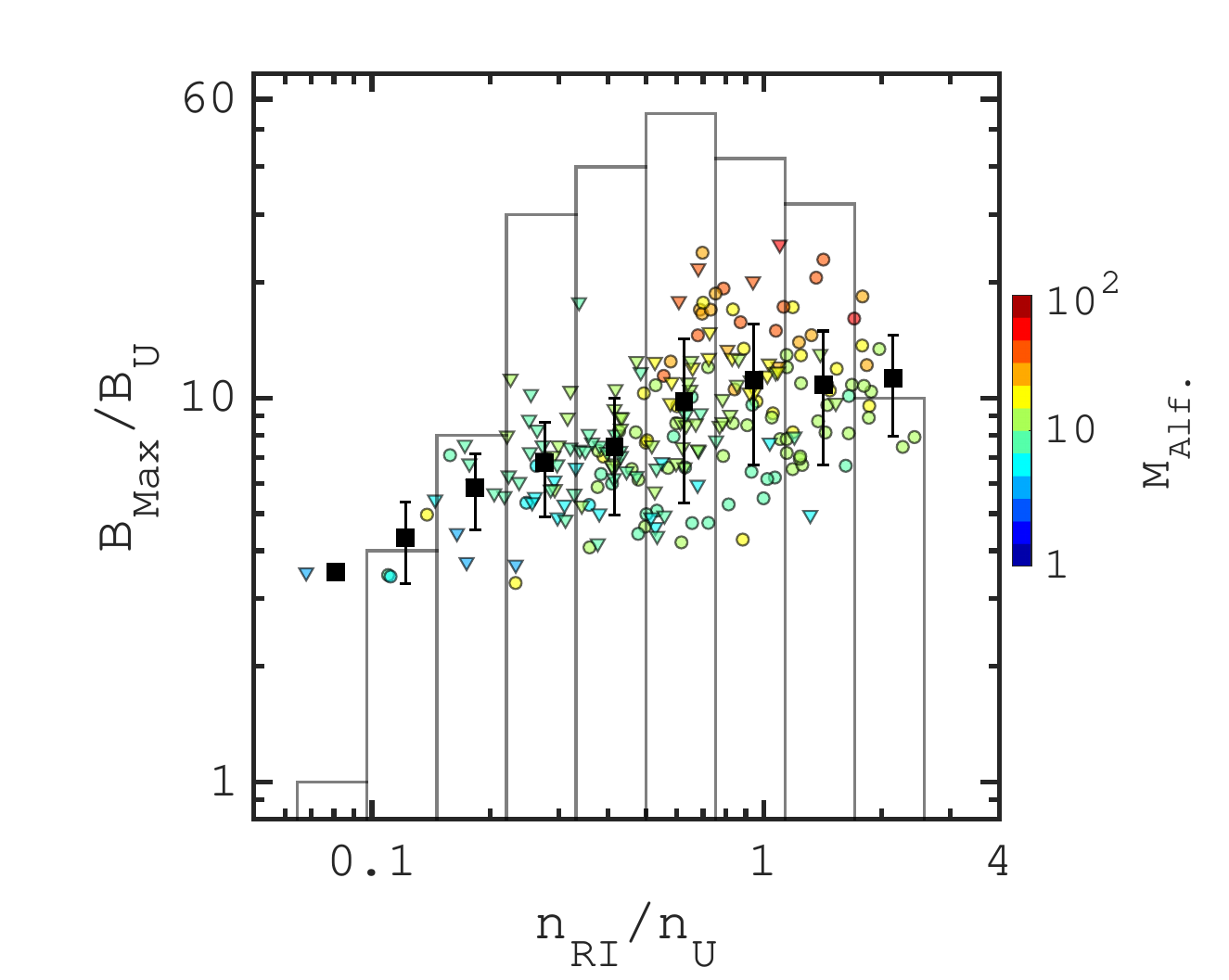}
\caption{\textbf{Magnetic amplification rate as a function of RI fraction.} The color scale represents the Alfvén Mach number of the shocks. The black squares are binned density averages with associated standard error within each bin shown with error bars. The histogram in the background shows the number of events in each bin. Data points are distinguished by upside-down triangle and circle markers for the analysis method based on ``RI Track'' and ``SW Subtract'', respectively. 
\label{fig:fig2}}
\end{figure}

We begin by quantifying the magnetic amplification rate across the shocks and by establishing a relationship between this parameter and the RI abundance using the large number of events in our dataset. In Figure \ref{fig:fig2} we compare the ion reflection ($n_{RI} / n_{U}$) and magnetic amplification ($B_{Max} / B_U$) rates normalized by corresponding upstream values ($B_U$ and $n_U$). Black squares on this figure are averaged binned data points showing the overall trend in the data. The magnetic amplification rate increases with ion reflection. For $n_{RI} / n_{U}$ between 0.2 to 0.5, $B_{Max} / B_U$ continues to increase but at a lower rate. We see further increase in the average magnetic amplification rate as the RI fraction increases beyond 0.6. But it remains constant at slightly above 10 for larger reflection rates. The plateau in the magnetic amplification rates suggests that with all other shock parameters held fixed, the increased in abundance of RIs cause magnetic perturbations across the shock up to a threshold reached at $B_{Max} / B_U \sim 10$. As the color scale in Figure \ref{fig:fig2} indicates, shock events that exhibit higher amplification rates also occur at higher Mach numbers (red colors).

\begin{figure}[hb!]
\centering
\includegraphics[width=0.7\textwidth]{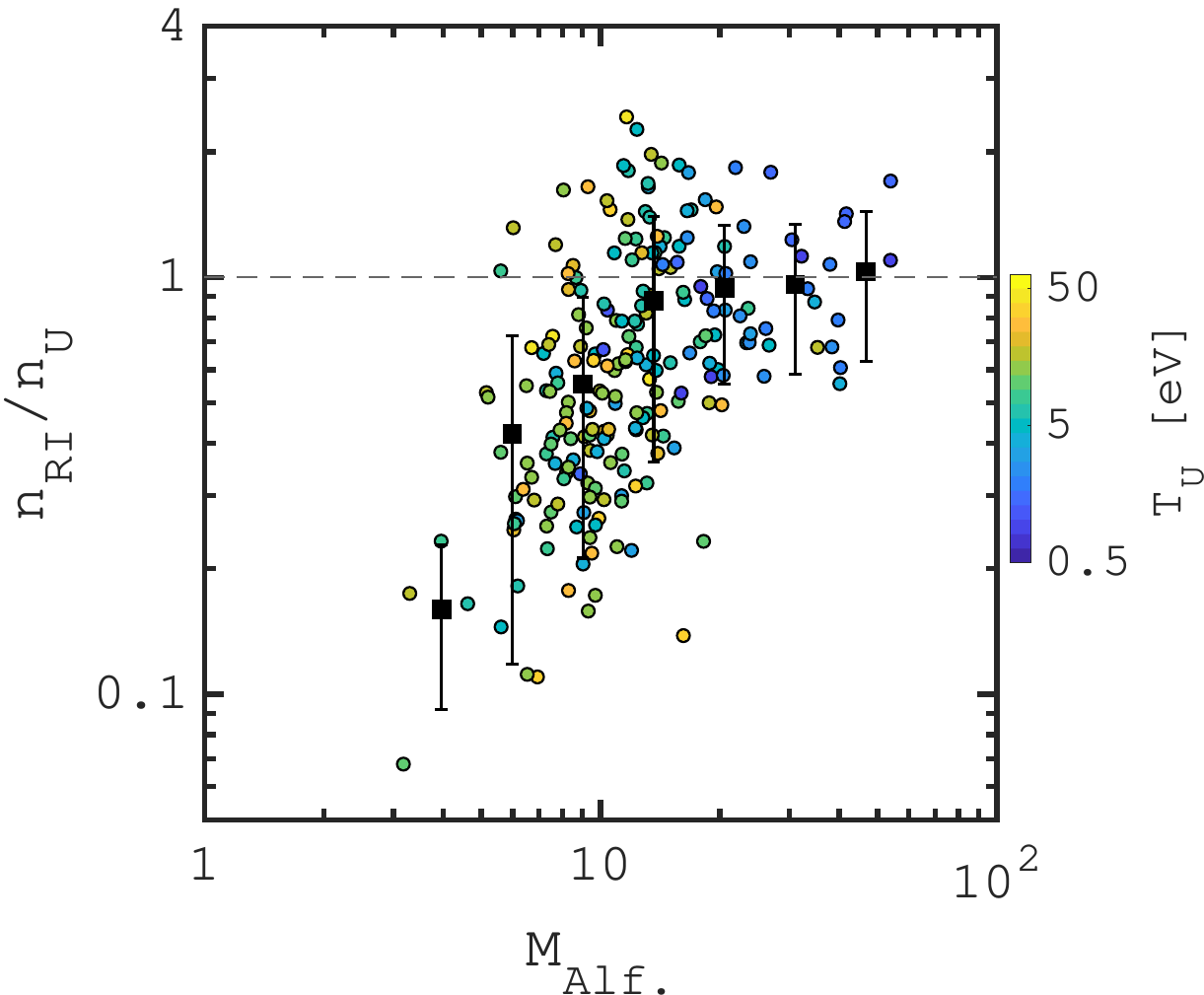}
\caption{\textbf{Ion reflection rate as a function of Alfvén Mach number.} Data are color-coded by the upstream plasma temperature. The black squares are averages of binned densities. The associated standard error within each bin is shown with error bars. The horizontal dashed line is drawn at $n_{RI} / n_{U} = 1$.}
\label{fig:fig3}
\end{figure}

In Figure \ref{fig:fig3} we show changes in the fraction of RIs as a function of $M_{Alf}$. Similar to Figure \ref{fig:fig2}, the black squares are averages of binned data and they show that ion reflection at $Q_{\perp}$ shocks increases with the upstream Mach number. In addition, with increasing upstream energy ($M_{Alf} > \sim 20$), shocks approach ``full reflection'' $n_{RI} / n_{U} \sim 1$, or when there are equal or even more ion densities reflected from the shock than there are ions in the far upstream and pristine solar wind. This effect is known as the precursor effect and is discussed in more details in Figure \ref{fig:sim_fig}. The $n_{RI} / n_{U} \sim 1$ threshold explains the flattening of the magnetic amplification rate observed in Figure \ref{fig:fig2}. The ion reflection rates shown in Figure \ref{fig:fig3} are much higher than the predictions of analytical models \citep{leroy_structure_1982, wilkinson_parametric_1990}. While such models also predict that the ion reflection rate increases with Mach number, they limit the reflection rate at $\sim 0.2$ times the upstream plasma density. The reflection rate in these models is highly sensitive to the adiabatic index chosen in the Rankine-Hugoniot relations. The constant reflection rate of 0.2, and in some cases values as high as 0.5, is routinely adopted in theoretical studies of kinetic effects and instabilities at collisionless shocks \citep{matsukiyo_modified_2003}. Changes to this parameter can modify the solution to the dispersion relation, the amount of cross field current, and estimates of the instability growth rates upstream of the shock.

\begin{figure}[h!]
\centering
\includegraphics[width=0.7\textwidth]{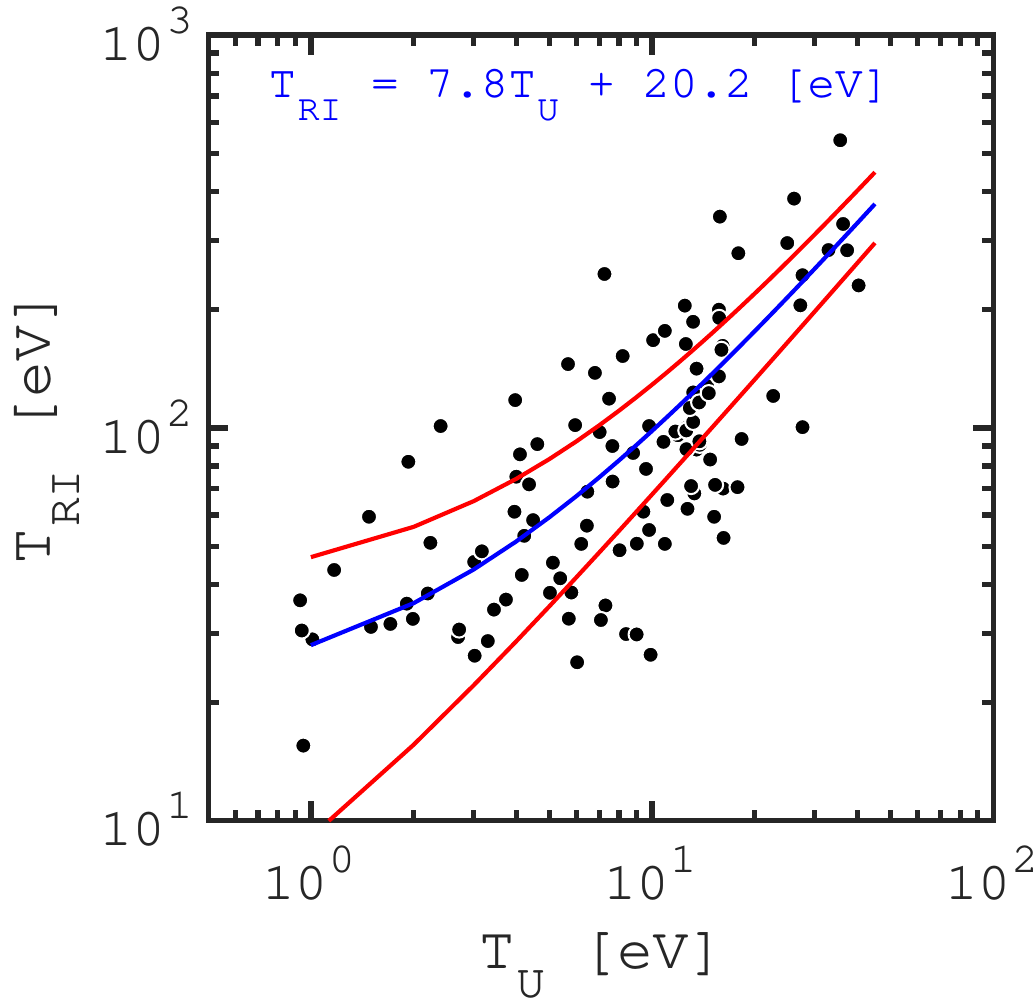}
\caption{\textbf{Reflected ion temperatures as a function of upstream ion plasma temperature.} Both axes are in units of electron-volt (eV). The linear fit (blue line) and the $95\%$ confidence intervals (red lines) are overplotted. The fitting function is annotated in blue on the top of the figure. RI temperatures can only be obtained for ``RI Track'' events.}
\label{fig:fig4}
\end{figure}

The data points in Figure \ref{fig:fig3} are color-coded by the upstream plasma temperature, indicating higher Mach number shocks are observed at colder solar wind plasmas. Some instabilities upstream of the shock are caused by temperature anisotropy caused by RIs that gyrate perpendicular to the magnetic field. We show the dependence of the RI beam temperature on the upstream plasma temperature in Figure \ref{fig:fig4}. RI temperatures are much higher than the upstream ion beam temperature. The first order linear fit shows the following relationship between the RI temperature and the upstream plasma temperature:
\begin{equation}
T_{RI} = 7.8 T_U + 20.0 [eV]
\end{equation}

\noindent The fit function and $95\%$ intervals are shown on Figure \ref{fig:fig4}. The R\textsuperscript{2} goodness-of-fit is 0.506. Several factors contribute to such high RI temperatures. Ion reflection is not necessarily always specular, and the spread of the reflected ion beam can change due to non-specular reflection. Especially at high Mach numbers, where upstream magnetic perturbations change the morphology of the shock surface \citep{krasnoselskikh_nonstationarity_2002}, some ions are non-specularly reflected or scattered upon reflection. In addition, non-local ion reflection and extended trajectory of RIs upstream of the shock further increases the velocity spread of RIs \citep{gedalin_non-locality_2023, madanian_transient_2023}.

While there seems to be a good correlation between the temperature of RIs and the upstream plasma temperature, we find no relationship between the RI temperature and the magnetic amplification rate. We see a strong correlation (not shown) between the magnetic amplification rates and the upstream Mach number, as has been shown in previous observational and analytical studies \citep{russell_overshoots_1982, sulaiman_quasiperpendicular_2015, fraschetti_turbulent_2013, bell_cosmic_2013}. In addition, magnetic amplification increases with the increase in the normalized drift speed of RIs $v_d=(|\textbf{\textit{V}}_U - \textbf{\textit{V}}_{RI}|)/ v_{Alf}$. These correlations lead to the fact that the magnetic field amplification is strongly driven by the ratio of upstream dynamic pressure to the magnetic pressure, a parameter also known as the inverse of the magnetization parameter of the upstream plasma ($1 / \sigma$,):

\begin{equation}
1 / \sigma = P_{Dyne.} / P_B = \frac{ \mu_{0} \rho V_{U}^2}{B_U^2}
\end{equation}

\begin{figure}[h!]
\centering
\includegraphics[width=0.7\textwidth]{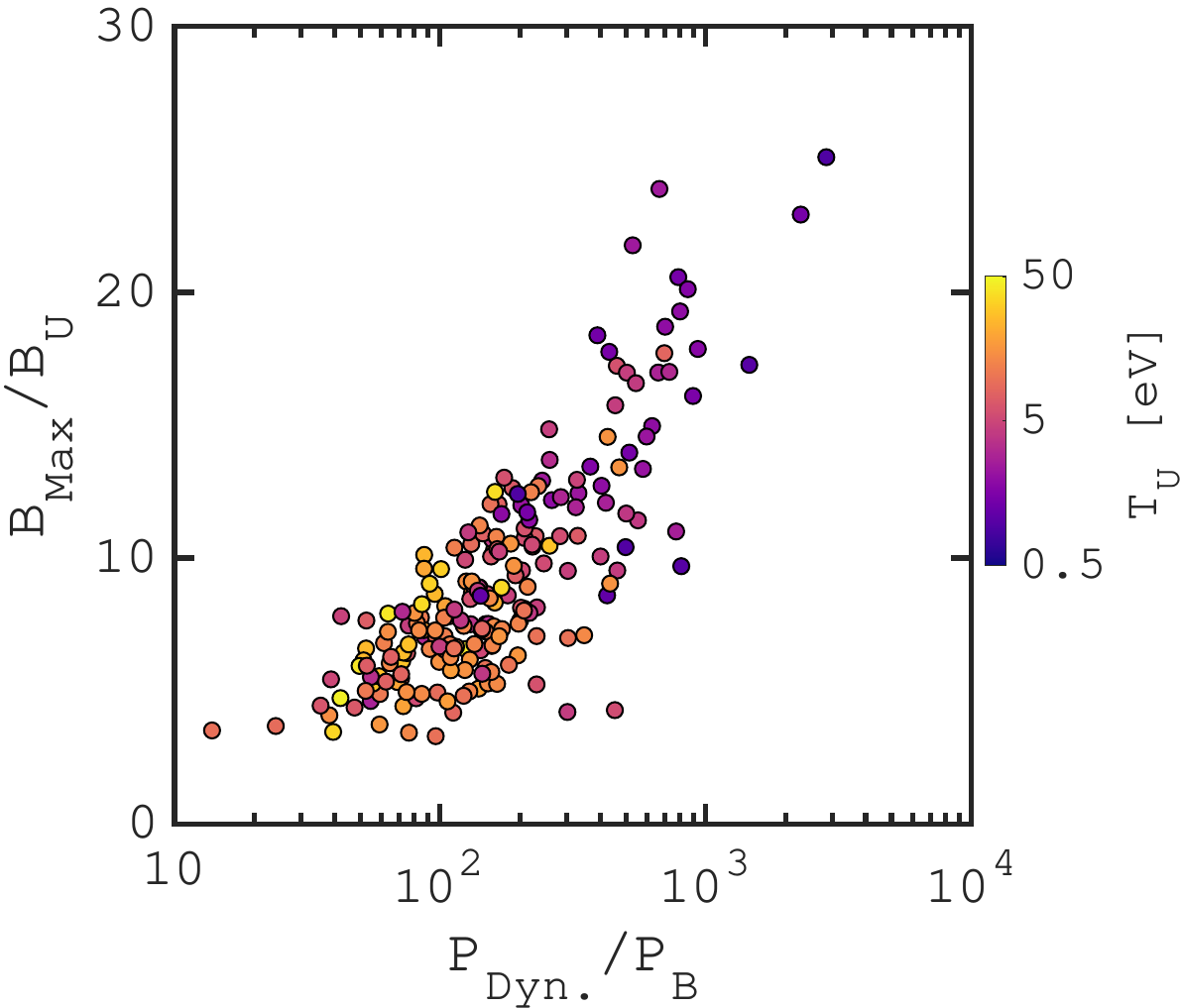}
\caption{\textbf{Magnetic amplification rate as a function of $P_{Dyne.} / P_B$ ($1 / \sigma$).} Data points are color-coded by the upstream plasma temperature. Events with low upstream plasma temperatures occur in weaker magnetic field strengths.}
\label{fig:fig6}
\end{figure}

The dependence of the magnetic amplification rate on inverse magnetization (or $M_{Alf}^2$) is shown in Figure \ref{fig:fig6}, where a strong positive correlation between the two parameters is observed.

Increase in the RI drift speed and decrease in magnetization result in a larger ion gyroradius and a larger foot region to be transversed by ions. With decreasing magnetization, RIs can propagate farther upstream of the shock. Therefore, instabilities that they generate have more time to develop and waves have time to steepen as they traverse the foot, much like upstream waves in the foreshock region of quasi-parallel shocks \citep{chen_solitary_2022, chen_solitary_2021}. Interestingly, theoretical studies of relativistic shocks have argued that for low upstream magnetization, the magnetic field perturbation and the flux of cosmic rays generated at $Q_{\perp}$ shocks can be derived using the same formalism as that used for quasi-parallel shocks \citep{bell_cosmic-ray_2018,sironi_maximum_2013}.


As shown in Figures \ref{fig:fig2} and \ref{fig:fig3}, some events exhibit significant ion reflection rates $n_{RI} / n_{U} > 1$. This may seem to violate the conservation of particle flux at the shock. To validate our methodology, we perform a series of particle-in-cell simulations to examine the density ratio of RIs \citep{winske_magnetic_1988}. The simulation conditions are listed in Table \ref{tab:sim_params}. The model discussed here combines a fully-kinetic particle-in-cell treatment of the ions with a charge-neutralizing, massless and isothermal electron fluid. The electromagnetic fields and particle moments are advanced using Maxwell's equations in the low-frequency limit using the 'current advance method-cyclic leapfrog' (CAM-CL) algorithm \citep{matthews_current_1994}. The code is otherwise adapted from the particle-in-cell code EPOCH \citep{arber_contemporary_2015}. We use a 2D grid of size $(N_x,N_y) = (2400,160)$ and resolution $\Delta x,y = 0.15d_i$, where $d_i$ is the upstream ion inertial length. The model is ``2.5D'' such that all three vector components electromagnetic fields and particle moments may vary on the two-dimensional grid, e.g. $B_{x,y,z}(x,y,t)$. The boundary conditions are periodic at the upper and lower edges in $y$-direction, reflecting at the downstream edge, and a source of inflowing solar wind plasma is generated at the upstream edge. The interaction of the flow with the reflecting boundary generates a shock moving in the $-x$ direction. The initial upstream conditions are uniform with normalized number density $n_0$, magnetic field $\mathbf{B} = B_U [ - \cos(\theta_{Bn}), \sin(\theta_{Bn}), 0]$, and solar wind inflow velocity $\mathbf{V} = [V_U,0,0]$.  The ions and electron fluid are initialized with plasma beta $\beta_i = 0.5$ and $\beta_e = 1.1$. In order to reduce the noise as far as possible, we sample the ion phase space with 200 pseudo-particles per cell.

We determine the density ratio of RIs by examining individual ion trajectories in the simulations. An ion is labeled as reflecting if it occupies the region of phase space defined by $(v_{i,x}-v_{sh})^2 + v_{i,z}^2 > v_{sh}^2$ while upstream of the shock ramp. The ion velocities $v_{i,x}$ and $v_{i,z}$ and shock velocity $v_{sh}$ are given in the downstream frame. We then find the density ratio by comparing the number of reflecting ions $n_{p,ref}$ to the number of inflowing ions $n_{p,SW}$. This analysis is performed for all times after $t = 5 \Omega_{i}^{-1}$ so that the shock structure is well-developed and unaffected by the downstream reflecting boundary.

The time-averaged density ratios as a function of distance from the shock ramp is shown in Figure \ref{fig:sim_fig}. In Panel (a), showing the U6T65 case, we observe an increase of the density of the inflowing ions up to $\sim 1 d_i$ upstream of the shock ramp. This is due to the solar wind beam slowing down (and therefore compressing) in the shock foot. We also note that, for this case, the density of reflected ions is higher than the density of inflowing ions for $x < 0.5d_i$, which is what we observe in MMS data for some shock crossings. In Panel (b) we compare the RI density ratios for all simulation runs. Generally, we observe higher densities of RIs for higher Mach numbers near the shock ramp, which supports our approach in selecting the representative RI density closest to the shock in time series data. In comparing runs U6T65 (orange) and U6T80 (brown) we find that the shock orientation $\theta_{Bn}$ does not strongly affect the density of RIs. We also note that at higher shock speeds and when $\theta_{Bn}$ approaches the quasi-parallel regime, a small fraction of RIs become back-streaming ions and escape the simulation box.

\begin{table}
\caption{\label{tab:sim_params} Shock parameters for the set of simulations described in this study, including inflow speed $V_U$, $M_Alf$ and $\theta_{Bn}.$}
\centering
\begin{tabular}{lccc}
\hline
ID & $V_U/v_{Alf}$  & $M_{Alf}$  & $\theta_{Bn} (^\circ)$\\
\hline
U3T65 & 3.0 &  3.2 & 65\\
U6T65 & 6.0 & 6.5 & 65\\
U9T65 & 9.0 & 9.7 & 65\\
U12T65 & 12.0 & 12.8  & 65\\
U6T50 & 6.0 &  6.5  & 50\\
U6T80 & 6.0 &  6.5  & 80\\
\hline
\end{tabular}
\end{table}

\begin{figure}
\centering
\includegraphics[width=\textwidth]{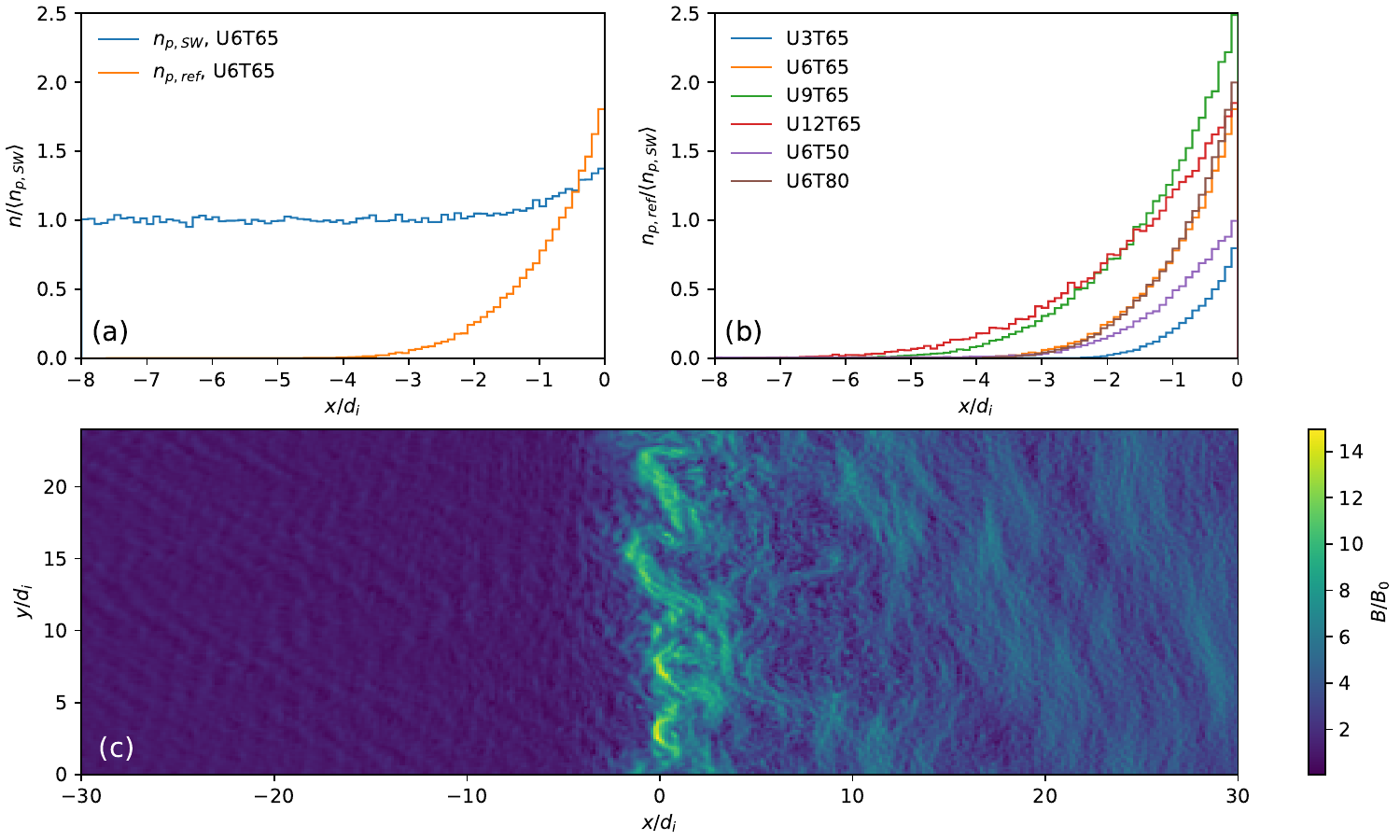}
\caption{
Simulation results for the density of reflected ions $n_{p,ref}$ normalized by the mean upstream solar wind density $\left<n_{p,SW}\right>$, as a function of distance $x$ upstream of the shock ramp. Panels are as follows: (a) Comparison of the density of reflected ions to the density of inflowing solar wind ions for the run U6T65, (b) Comparison of the normalized density of reflected ions for all of the simulations, and (c) Magnetic field strength in the shock layer for run U6T65 at a single timestep, showing upstream enhancements and non-planar shock morphology.}
\label{fig:sim_fig}
\end{figure}

\section{Discussions and Conclusion}

Magnetic amplification at quasi-perpendicular shocks is an energy conversion problem. We address factors involved in field amplification from an ion kinetic perspective. We use in-situ spacecraft observations during 222 crossings of the Earth’s $Q_{\perp}$ bow shock to quantify the ion reflection and magnetic amplification over a wide range of upstream conditions. The upstream Alfvén Mach number in these events varies between 3.1 and 53.9, the upstream plasma temperature is in the 0.9 – 45.1 eV range, the upstream flow speed relative to the shock varies between 296.0 and 748.8 kms\textsuperscript{-1}, the upstream magnetic field strengths range between 0.8 and 12.6 nT, while the magnetization varies between $3.51 \times 10^{-4}$ and 0.07 across all events.

Increase in the abundance of RIs contributes to magnetic amplification up to a certain limit. Both magnetic amplification and ion reflection increase with upstream Mach number till the full reflection condition is reached. We find that the magnetic amplification rates are closely correlated with RI drift speed, and decreasing magnetization in the upstream plasma. All these factors result in a larger ion gyroradius and creating a larger foot region upstream of a $Q_{\perp}$ shock. Ions can propagate farther upstream as the magnetization decreases to create magnetic perturbations that are steepended across the foot and contribute to higher magnetic amplification. 

The density of RIs upstream of $Q_{\perp}$ shocks can be substantially higher than previous analytical estimates. Our simulations in Figure \ref{fig:sim_fig} indicates that very close to the shock layer, the solar wind beam itself experiences signs of compression and increased beam density from the pristine conditions. In other words, the nature of the upstream conditions (Mach number, plasma beta, etc.) just upstream of the shock ramp intermittently changes. While non-negligible momentum and current density are carried by RIs away from the shock which can induce a response in the incident solar wind beam, waves and magnetic structures that develop upstream of the shock, as shown in Panel (c), interact with the incident ions and tend to create momentary compression in the flow close to the shock ramp. 

This precursor effect (not to be confused with the whistler precursor) is likely transient and localized within the region of intense ion reflection, and will evolve as the ion reflection rate adjusts to the new conditions. The effect can be viewed as the shock’s attempt to intermittently change the local upstream plasma in order to process the upstream energy; or as an extended shock transition region beyond the shock ramp where plasma heating lags compression, and heating of the solar wind is mainly through dissipation of ions rather than resistive processes. Such an understanding of shocks has not been realized in previous spacecraft observations, as the required time resolution to detect this effect has only become available recently through MMS observations. We see enhanced densities of RIs ($n_{RI} / n_{U} > 1$) in both types of events, ``RI Track'' and ``SW Subtract'', though there are more events under the latter category. Because of the way ``SW Subtract'' events are analyzed, RI densities under this category can still have some contamination from other populations.

The upper limit of the Mach numbers reported here is close to Mach numbers expected at supernova remnant shocks ($M_{Alf} \sim 50$) \citep{caprioli_cosmic-ray-induced_2013}. The highest Mach number shocks in our analysis are observed during upstream periods of very low magnetic field strength carried in a cold plasma flow (i.e., low Alfvénic and sonic speeds). This indicates that to study magnetic amplification at higher Mach numbers, spacecraft observations of planetary bow shocks at larger heliocentric distances must be utilized. The positive correlation between the interplanetary magnetic field strength and the proton temperature is a characteristic of the expanding solar wind plasma experiencing adiabatic cooling. As such, interpretation of changes in derived parameters based simultaneously on the upstream magnetic field strength and the ion temperature, such as plasma $\beta$, should be taken with caution. A highly dense, cold, and weakly magnetized solar wind plasma can have a $\beta$ value similar to a hot and moderately magnetized solar wind, though the shock profile will be vastly different between the two cases. 

There are only a few low Mach number shocks in our dataset despite such shocks being readily accessible at the Earth's bow shock. Although theoretically shocks become supercritical and begin reflecting ions to dissipate energy at $M_{Alf} \sim 2.7$ and above \citep{Marshal1955}, ion reflection at such low Mach numbers is rather sparse and difficult to detect even in high resolution data. Further, for $M_{Alf}$ below the critical whistler Mach number, $M_{wh}=1/2 \sqrt{m_p / m_e} \cdot cos(\theta_{Bn})$ where $m_e$ is the electron mass, shock generated whistler waves can phase stand upstream of the shock, interact with and scatter both the solar wind beam ions and RIs \citep{krasnoselskikh_nonstationarity_2002}. This makes it difficult to isolate either populations.

We finally emphasize that such high magnetic amplification rates reported in this study are not typically expected at $Q_{\perp}$ shock observations, and is hardly achieved in the numerical shock analysis. For instance, in 2D PIC simulations of a perpendicular shock with upstream magnetization of $6 \times 10^{-5}$ ($M_{Alf} \sim 130$), a magnetic amplification of only a factor of 15 was achieved \citep{kato_nonrelativistic_2010}. In-situ observations reported here provide a quantitative measure of the relationship between ion reflection and magnetic field enhancement rates under various upstream conditions. Our observations show that magnetic amplification at $Q_{\perp}$ shocks propagating in weakly magnetized plasmas is comparable to those at quasi-parallel shocks.

\begin{acknowledgments}
We thank Y. Khotyaintsev for useful discussions. This work was supported in part by the MMS project Early Career Program through the National Aeronautics and Space Administration (NASA) grant number 80NSSC23K0009. I. Gingell is supported by the Royal Society University Research Fellowship URF\textbackslash R1\textbackslash 191547.  
\end{acknowledgments}

\software{Data and Code Availability}

All data used in this study are hosted by NASA and are publicly accessible at \url{https://spdf.gsfc.nasa.gov/pub/data/mms}. 
The Space Physics Environment Data Analysis Software (SPEDAS) used in this work is publicly available at \url{http://spedas.org}

\appendix

\section{Supplementary Figures}
\begin{figure}[hb]
\centering
\includegraphics[width=0.7\textwidth]{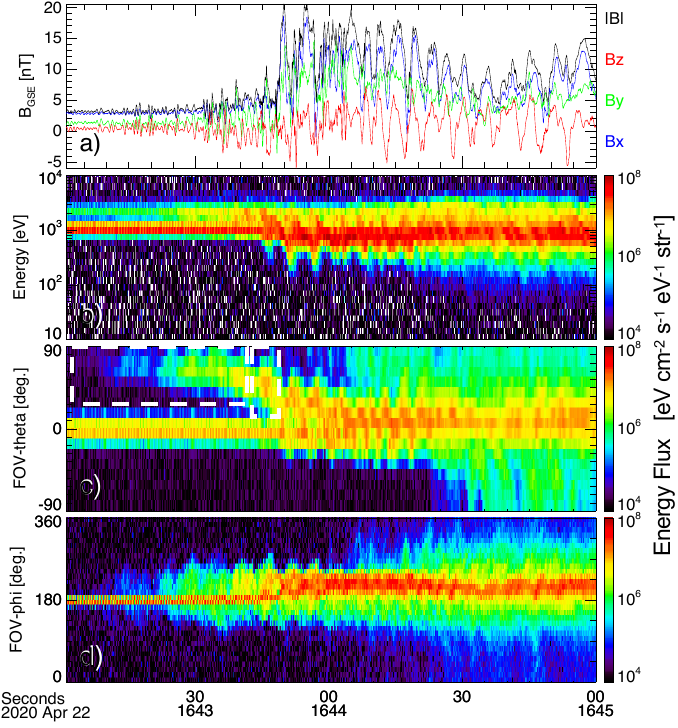}
\caption{Examples of shock crossings for ``RI Track'' (left) and ``SW Subtract'' (right) analysis methods. An example of spacecraft measurements of a quasi-perpendicular shock for which we track the reflected ions (``RI Track'') in the FPI FOV. Panels show time series data of: a) magnetic field strength and components in GSE coordinates, b) ion energy flux spectrogram, c) ion energy flux in the polar angle ($\theta$) in the FPI FOV, and d) ion energy flux along the azimuth $\phi$. The shock layer is at 16:43:55 UT. On panel c we mark the FOV angles associated with RIs with white rectangles, which are used in calculation of partial moments. The solar wind beam is evident at $\theta \sim 0^{\circ}$.}
\label{fig:supp_ref_track}
\end{figure}

\begin{figure}
\centering
\includegraphics[width=0.7\textwidth]{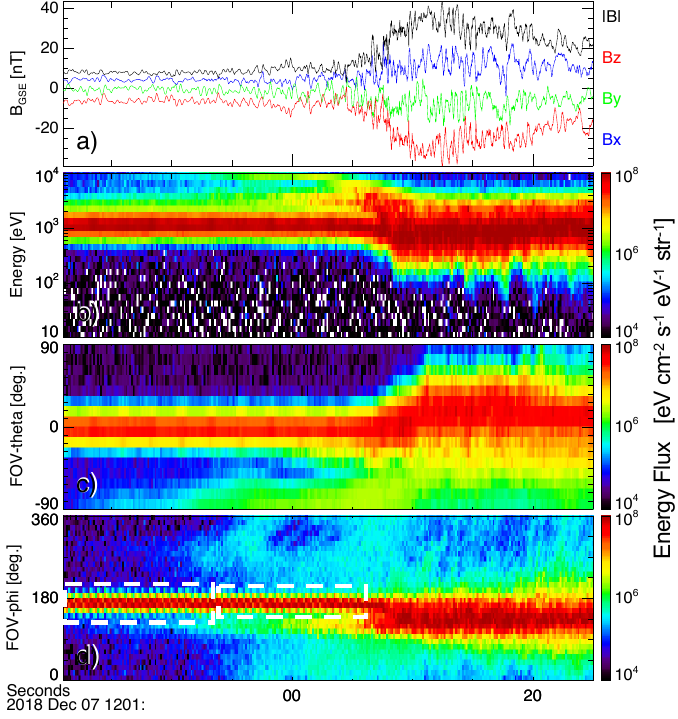}
\caption{An example of spacecraft measurements of a quasi-perpendicular shock for which we track the solar wind beam and subtract it from the total ion density to obtain the reflected ion densities (``SW Subtract''). The format of the figure is similar to \ref{fig:supp_ref_track}. The shock layer is at 12:01:06 UT for this event. A substantial overlap between the solar wind ion beam and non-solar wind ions is evident in panels c and d. On panel d we mark the angles associated with the solar wind beam with white dashed rectangles, which are used in calculation of partial moments.}
\label{fig:supp_sw_subt}
\end{figure}

Examples of shock crossings for each analysis type are shown in Figures \ref{fig:supp_ref_track} and \ref{fig:supp_sw_subt} in this section.

\end{document}